# scientific reports

**OPEN**

# Tunable circular dichroism through absorption in coupled optical modes of twisted triskelia nanostructures

Javier Rodríguez-Álvarez[1,2]✉, Antonio García-Martín[3], Arantxa Fraile Rodríguez[1,2], Xavier Batlle[1,2] & Amílcar Labarta[1,2]

We present a system consisting of two stacked chiral plasmonic nanoelements, so-called triskelia, that exhibits a high degree of circular dichroism. The optical modes arising from the interactions between the two elements are the main responsible for the dichroic signal. Their excitation in the absorption cross section is favored when the circular polarization of the light is opposite to the helicity of the system, so that an intense near-field distribution with 3D character is excited between the two triskelia, which in turn causes the dichroic response. Therefore, the stacking, in itself, provides a simple way to tune both the value of the circular dichroism, up to 60%, and its spectral distribution in the visible and near infrared range. We show how these interaction-driven modes can be controlled by finely tuning the distance and the relative twist angle between the triskelia, yielding maximum values of the dichroism at 20° and 100° for left- and right-handed circularly polarized light, respectively. Despite the three-fold symmetry of the elements, these two situations are not completely equivalent since the interplay between the handedness of the stack and the chirality of each single element breaks the symmetry between clockwise and anticlockwise rotation angles around 0°. This reveals the occurrence of clear helicity-dependent resonances. The proposed structure can be thus finely tuned to tailor the dichroic signal for applications at will, such as highly efficient helicity-sensitive surface spectroscopies or single-photon polarization detectors, among others.

Chiral structures play a key role in the working mechanism of a wide variety of biological processes and biochemical interactions, and thus hold promise for several technological applications. The term "chiral" refers to structures which are not superimposable with their mirror image. Therefore, any chiral structure can be found in two different handedness formed of the same building blocks. One of the most relevant properties of chiral structures is that the two versions of the system react differently under the illumination of left-handed (LCP) and right-handed (RCP) circularly polarized light. This chiroptical activity has been extensively used in many fields, such as chemistry[1,2], pharmaceuticals[3], and optics[4,5]. For a given chiroptical system, the differences in the signal of the optical functions $f$ of the system under the two circular polarizations is usually named circular dichroism (CD) and can be quantified by the dimensionless figure of merit (FOM)

$$CD_f = \frac{f_{LCP} - f_{RCP}}{f_{LCP} + f_{RCP}},$$ (1)

where $f$ stands for any of the studied optical cross-sections (CS), namely, the absorption, scattering, and the extinction, recorded under either LCP or RCP light. Although there are other alternatives, Eq. (1) is a common definition in the literature[6,7] to ascertain the significance of any dichroism present in the spectra of the optical functions of a system.

Due to the versatility of plasmonic metamaterials, several plasmonic chiral systems have been proposed in the last years in response to a growing interest in the experimental realization of such chiral nanostructures. The approaches to their manufacture comprehend the realization of three-dimensional (3D) structures[8], the manipulation of nanoparticle assemblies[9,10] or the stacking of pseudo-planar structures[11,12], among others. Even though planar structures can be easily manufactured by standard lithography processes[13], in the recent years, 3D plasmonic structures have gained interest because of their enhanced chiral response and simple tunability, compared to their 2D counterparts[14–16]. Despite the good performance of 3D structures, their manufacture can be highly demanding and limited by practically attainable feature sizes and complex manufacturing procedures, especially if a strong CD in the visible range is desired. To address this issue, stacked 2D nanostructures made of simple elements following a multilayered design can fill in the gap as they have shown to exhibit broadband dichroic signals[12,17–24]. Although the twice exposure of the pattern with high alignment precision during nanofabrication makes stacked 2D nanostructures also challenging, they might be easier to manufacture when based on simple designs than 3D structures. Thus, stacked 2D structures can be broadly found in many applications related to nanophotonics[25–27]. Despite their apparent simple design, complex interactions among the plasmonic nanoelements acting as basic building blocks of these structures enable to manipulate light in a highly efficient way. For instance, the optical response of the system can be tuned by simply controlling the spatial arrangement of the building blocks in the structure.

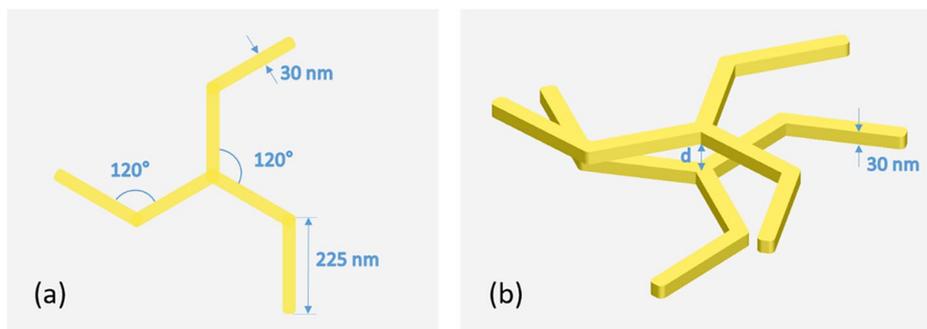

**Figure 1.** (**a**) Schematic top view of one triskelion showing the values of the main in-plane geometrical parameters. The thickness of the triskelion was set to 30 nm. (**b**) 3D-view showing the parallel stacking of the two triskelia in the system. The edge-to-edge distance $d$ between the two triskelia is also depicted. Note that the triskelion on top of the stack is twisted 30° anticlockwise with respect to the bottom one.

Unlike previous realizations of stacked structures[12,28], the plasmonic nanoelements used as building block in this work has a handedness such that they are optically active in nature. This yields to the arising of additional resonances driven by the interaction between the two elements. Consequently, the CD of the structure can be controlled by the in-plane relative rotation angle (twist angle) between the basic elements with respect to the handedness of the element itself together with the distance between the elements. We therefore present a system consisting of two stacked chiral plasmonic nanoelements with three-fold rotational symmetry, so-called triskelia, providing a high degree of CD in the visible to near-infrared range that could be easily tuned in a manufacturing process.

## System and numerical methods
### Numerical simulations.
The simulations presented hereafter have been performed using a commercial Finite Difference Time Domain (FDTD) method providing a robust and reliable solver for Maxwell's equations (Lumerical[29]). We use an impinging plane wave along the $z$-axis (perpendicular to the surface of the triskelia) of unit amplitude everywhere. The simulation cell is large enough ($1.2 \, \mu m \times 1.2 \, \mu m \times 4 \, \mu m$) to ensure that perfectly absorbing boundary conditions exert a negligible effect on the electromagnetic fields obtained. We employ a parallelepiped mesh in the nanostructures and near field region of $1.5 \, nm \times 1.5 \, nm \times 1 \, nm$, $dx$-$dy$-$dz$, respectively, growing uniformly in $dz$ up to a maximum of 25 nm out of the nearfield close to the simulation boundaries, so that convergence (to the best of our numerical capabilities) is attained. The total and scattered fields are then collected to give rise to the nearfield intensity color maps and the cross-sections. All fields are normalized to the amplitude of the impinging plane wave.

### Design of the triskelia stacking.
Prior to the design of more complex chiral structures, it is convenient to choose a planar monomer showing some dichroism in its optical response that can be appropriate as building block of 3D chiral systems. We have chosen a monomer made up of three elements emanating from a center with three-fold rotational symmetry so that its electric polarization is not fully compatible with an even number of poles[30,31]. This introduces a certain geometric frustration that favors the formation of more complex polar distributions, which in turn may make the structure more prone to show dichroism under circularly polarized illuminations. In addition, the three elements of the monomer are bent at their midpoint forming the same clockwise angle so that the three-fold rotational symmetry is preserved while not presenting any mirror plane parallel to the rotation axis. Figure 1a depicts the design of the monomer used in this work as the primary source of planar CD. It has been named "triskelion" in view of its resemblance to the artistic motif and the three-legged clathrin molecule[32]. Interestingly, any planar structure embedded in a homogeneous dielectric medium, even fulfilling the stated properties, is not truly chiral because of reciprocity and the existence of an inherent symmetry mirror plane[33–35], and although significant CD signals can be exhibited by both the absorption and the scattering CS, they cancel out when the total extinction is considered. One way to induce true 3D chirality in a planar structure is just to put it on a dielectric substrate, so that both sides of the structure are not equivalent[35,36]. Nevertheless, in this work we propose a 3D arrangement of two of these planar motifs in a homogeneous medium that consists of a stack of two parallel interacting triskelia separated by an edge-to-edge distance $d$ (Fig. 1b) aiming to improve both the tunability and the intensity of the dichroic signal. This design is somehow similar to those in Refs.[28,37], where the monomers have no in-plane mirror symmetry, but here we take advantage as well from the interactions between the electric polarizations of monomers with three-fold rotational symmetry. Additionally, the parallel stacking provides a simple platform to study the CD of the system as a function of both the edge-to-edge distance between the two triskelia and the twist angle $\varphi$ with respect to each other (the twist angle refers to the relative in-plane rotation between elements). It is worth stressing that these two geometric parameters could be easily tuned in a manufacture process. The material chosen for the study of the optical response of the system is gold owing to its well-known plasmonic properties and chemical inertness[38,39]. Numerical simulations of the absorption and scattering CS for a single triskelion with 30 nm in thickness show that the corresponding CD signals are maximized (not shown) when all the angles in Fig. 1a are kept equal to 120°.

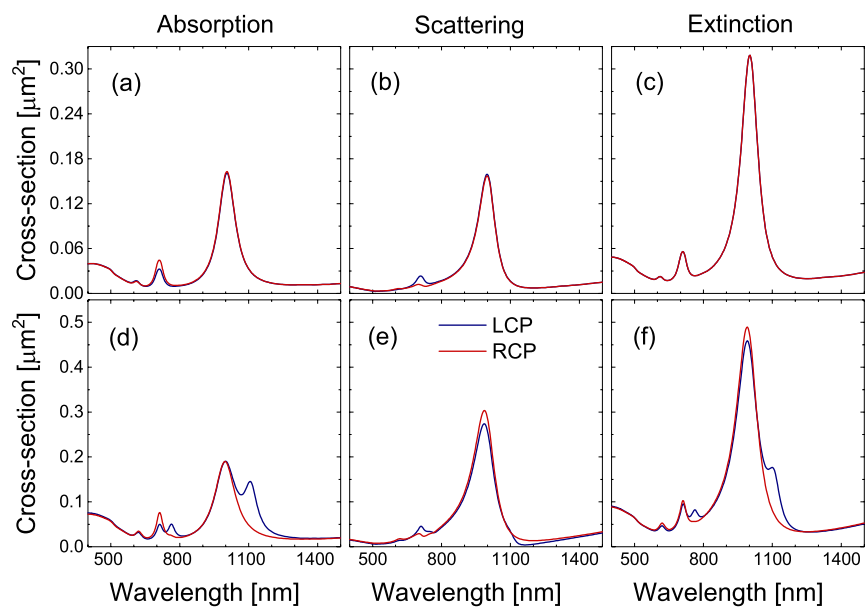

**Figure 2.** (**a**) Absorption, (**b**) scattering, and (**c**) extinction cross-sections under LCP (blue solid line) and RCP (red solid line) light for a single triskelion (see Fig. 1a). (**d**) Absorption, (**e**) scattering, and (**f**) extinction CS under LCP (blue solid line) and RCP (red solid line) light for a double triskelion system forming an anticlockwise 30° twist angle and at an edge-to-edge distance $d$ = 30 nm (see Fig. 1b).

## Results and discussion

**Optical responses of the single triskelion *vs* the stacking.** A priori, the planar nature of a single triskelion in combination with reciprocity[33,35] is known to prevent any CD in the total optical extinction. However, as shown in Fig. 2a,b, neither the absorption nor the scattering holds this suppression by themselves. A close look at Fig. 2a,b reveals different signals for RCP and LCP light both in the absorption and scattering channels, predominantly around the peak at 710 nm, whereas the CD perfectly vanishes in the extinction CS (Fig. 2c) since a planar triskelion embedded in a uniform dielectric medium is not a truly chiral structure[33]. However, when triskelia are set to form a non-planar structure, the optical properties of the whole system are expected to change drastically, enabling the existence of CD in the extinction CS.

As an example, Fig. 2d,e show the absorption and scattering CS for a triskelion stack with and edge-to-edge distance of 30 nm and anticlockwise 30° twist angle between the top and bottom triskelia (see Fig. 1b). Similar peaks around 710 and 1000 nm than those for a single triskelion can be identified in the absorption and scattering CS under both RCP and LCP illuminations. Nevertheless, the absorption CS under LCP light also exhibits two extra peaks, which are redshifted with respect to those around 710 and 1000 nm and are the main cause of CD in the extinction CS since they have no counterparts in any of the scattering CS for the two light polarizations (see Fig. 2d–f). Therefore, the interaction between the electric polarizations of the two triskelia under LCP illumination is causing these two extra excitation modes in the absorption CS that are in turn responsible for the CD in the extinction CS.

**Tuning interactions: relative in-plane angle between the triskelia.** As shown earlier in Fig. 2d–f, the optical response of the system can be tuned by changing the twist angle $\varphi$ between the two triskelia in the stack. This will effectively modify the distance between the legs of the two triskelia, as well as induce a "sense of turn"[13,35]. In fact, when $\varphi = 0$ (or an integer multiple of 120°, as a single triskelion has three-fold rotational symmetry), from the point of view of the CD the system can be regarded as equivalent to the planar case of a single element irrespective of the edge-to-edge distance between the two triskelia since no additional helicity is associated with the 3D stacking, as also shown in Ref.[37]. Therefore, for those twist angles, the system only displays some differences in the absorption and scattering CS mostly around the peak at 710 nm and smaller around 1000 nm but not in the extinction CS as in the case of a single triskelion (see Fig. S1 in the Supplementary Information) since this parallel stack does not modify the planar nature of each triskelion. Thus, the CS are very much like those of a single triskelion in Fig. 2a,b, but with some important distinctions. First, the scattering CS is almost twice as intense as for the single triskelion, whereas the absorption CS remains equally intense. Second, there are two extra peaks (resonances) in the absorption CS caused by the interaction between the two triskelia, one barely noticeable around 850 nm and another one more visible around 1250 nm that appear under both LCP and RCP light with the same intensities so that they do not contribute to the CD in agreement with the no 3D helicity of the stack for this special value of the angle. As will be discussed later, the spectral positions of these two extra peaks depend on the distance between the triskelia as they are driven by interactions.

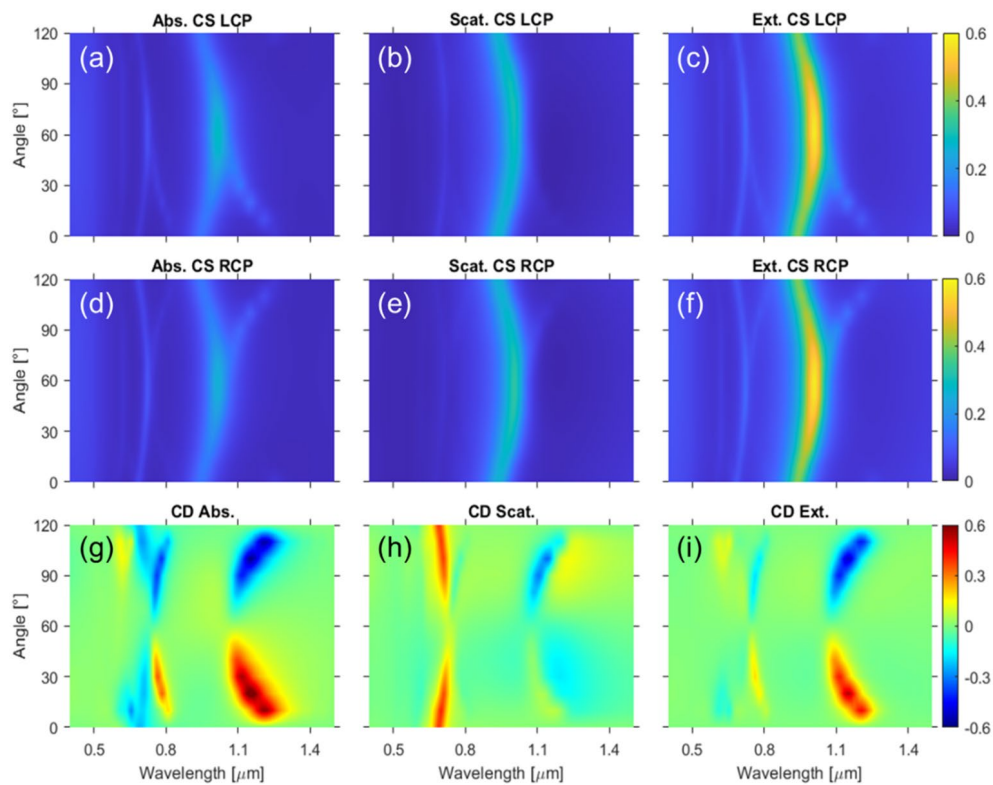

**Figure 3.** Absorption, scattering, and extinction CS for LCP (**a–c**), and RCP (**d–f**), illuminations, respectively, as a function of the twist angle $\varphi$ for a fixed edge-to-edge distance of 30 nm. FOM defined in Eq. (1) for the CD of the absorption (**g**), scattering (**h**), and extinction (**i**) as a function of the twist angle.

Figure 3a–f shows the absorption, scattering, and extinction CS, for both LCP and RCP incidence, as a function of anticlockwise $\varphi$ (see Fig. 1), for an edge-to-edge distance of 30 nm, together with their respective FOM for the CD of the three CS (Fig. 3g–i). By varying $\varphi$, the spectra of all CS display the two main excitations at ca. 710 nm and 1000 nm (already shown by a single triskelion) that are always present under both circular polarizations and, more importantly, the emergence of two extra excitations with intensities which vary greatly as a function of the handedness of the circular polarization. For instance, it is found that the extra excitations in the absorption, which are clearly visible under LCP incidence at an anticlockwise $\varphi \leq 60°$ (see Fig. 3a and Fig. 2d for the case of 30°), are then predominant under RCP illumination for $60° < \varphi < 120°$ anticlockwise (see Fig. 3d and Fig. S1 in the Supplementary Information for $\varphi = 90°$). Maximum values for the CD in the extinction CS can be found for $\varphi = 20°$ (and $\varphi = 100°$) for the excitation around 1000 nm, reaching CD values up to 60%. High values of the CD can also be found in a broad range of parameters around the maximal value. In addition, remarkable CD values are also shown at the resonance in the vicinity of 710 nm. For twist angles of 30° and 90° we can observe a sharp increase in the CD. These results show that our structure presents values of the CD perfectly comparable to those of the state of the art, even outperforming similar works[7,40].

Since the range of anticlockwise twist angles within 0° and 60° is *almost* optically equivalent to that of the clockwise ones starting from 120°[12,41,42] (for instance, the optical response for 110° clockwise should be equivalent to that of 10° anticlockwise), these results put forward the occurrence of clear helicity-dependent excitations (chiral resonances) that are manifested as enhanced absorption peaks. We would like to point out the interesting features associated with two special angles. These are the aforementioned highly symmetric case of $\varphi = 0°$ that shows no CD (despite being driven by near-field interactions) and that of $\varphi = 60°$, which shows a peculiar behavior since the only element that breaks perfect symmetry is the bending angle at the midpoint of the legs of each triskelion.

Interesting enough is the case of $\varphi = 60°$, where the CD in the extinction CS vanishes (see Fig. S1 in the Supplementary Information and Fig. 3g–i), similarly to Ref.[20]. This angle represents a highly symmetric arrangement of the triskelia in the stack, for which symmetric dispersions (with opposite sign) of the main excitations in the absorption and the scattering CS occur when $\varphi$ is increased or decreased from 60°. These symmetric dispersions of the absorption (Fig. 3a,d) and the scattering (Fig. 3b,e) are equal under LCP and RCP, while only some differences are exhibited by the intensities of the peaks for the two polarizations. However, those differences yield CD signals almost canceling each other out in the extinction CS (Fig. 3i) because of the opposite sign of the two contributions. In fact, the only noticeable differences between the CS for the triskelion stack at $\varphi = 60°$ and those for the single triskelion are the intensities of the signals since the two extra excitations are not present in the spectra of the latter (see Fig. S1 in the Supplementary Information). It is also worth noting that 60° is the angle for which the distance between equivalent points in the legs of the two triskelia is maximum for a fixed value of the edge-to-edge distance such that the interaction among the legs of the two triskelia in the stack is minimum. Besides, this distance is equal for both clockwise and anticlockwise screw rotation through the stack, so no dependence of the excitation on the handedness of the light can be expected.

The incomplete complementarity of the double triskelion structure comes from the bending angle at the midpoints of the legs of each triskelion. Perfect complementarity would be achieved if the sign of the bending angle of the legs were fully reversed[12], and thus LCP and RCP spectra could be interchanged for anticlockwise and clockwise in-plane angles within 0–60° and 120–60°, respectively. In terms of the CS spectra, the incompleteness arises from the fact that the extra excitations are also discernible in the scattering CS (Fig. 3b,e) for twist angles within 120–60°, although the contribution of the absorption to the CD in the extinction is still much more prominent. A clear example of the lack of perfect complementarity is set by the comparison of the scattering CS for 30° and 90° shown in Fig. 2d and Fig. S1 in the Supplementary Information, respectively. One of the extra excitations is perfectly discernible on the right-hand side of the peak around 1000 nm for 90° whereas it is not present at all for 30°.

Regarding the spectral dependence of the excitations on $\varphi$, the peaks arising from the single triskelion (denoted as main peaks) and those owing to the interactions between the two triskelia in the stack (so-called extra excitations) must be analyzed separately. Both kinds of excitations, exhibit spectral shifts of the corresponding CS as a function of $\varphi$, as expected for interacting systems (Fig. 3a–f). The main peaks are redshifted for $0° < \varphi < 60°$ or for $60° < \varphi < 120°$ owing to the symmetry around the 60° case. In contrast, the extra peaks evolve in a complementary fashion: they are blueshifted for angles $0° < \varphi < 60°$ or $60° < \varphi < 120°$.

Concerning the CD spectra, most of the interest lies in the contributions from the extra excitations. Despite being non-negligible for the main peaks, remarkable values are only obtained for the extra ones. Increasing the twist angle from 0°, for which the extra peaks have the same intensities under LCP and RCP illuminations and there is no contribution to the CD in the extinction cross-section (Fig. S1 in the Supplementary Information), the extra peaks under RCP light rapidly decrease and vanish above about $\varphi = 10°$, while their intensities monotonously increase for LCP light (see Fig. S1 in the Supplementary Information for $\varphi = 10°$ and Fig. 3 for 30°). An almost complementary behavior of these extra peaks under the two circular polarizations is found within 120° and 60° (see Fig. 3a–f). The contributions of the extra peaks to the CD spectra are redshifted from the highly symmetric angle of 60°, having opposite sign depending on the circular polarization of the light at which the corresponding extra excitations appear, as depicted in Fig. 3g–i. It must be pointed out that the high values of the CD in the scattering CS in the long wavelength range are mostly due to the intrinsically low values of the scattering CS for both polarizations.

### Tuning interactions: edge-to-edge distance.

As mentioned above, the other important geometrical parameter influencing the mutual interactions between the triskelia is the vertical separation (edge-to-edge distance) between them. In Sect. "Optical responses of the single triskelion *vs* the stacking", it was mentioned that the interaction between the two triskelia in the stack causes the emergence of extra peaks in the spectra of the CS. Some are only visible under the proper circular polarization of the impinging light. In Fig. 4, we show the absorption, scattering, and extinction CS for both LCP and RCP light, and the corresponding FOM for the CD, as a function of the edge-to-edge distance between the two triskelia upon a fixed anticlockwise twist angle $\varphi = 30°$. For this value, the extra peaks in the absorption CS are solely present under LCP illumination (Fig. 4a, d), as expected from the prior discussion in Sect. "Optical responses of the single triskelion *vs* the stacking". These extra excitations are blueshifted while the main excitations are redshifted as the distanceOptical responses of the single triskelion *vs* the stacking between the triskelia increases, tending to overlap and resulting in only two peaks as the interaction between the two triskelia decreases. Therefore, at a sufficiently large edge-to-edge distance the positions of the peaks coincide with those of a single triskelion but almost doubling the intensities. In contrast, the scattering CS does not exhibit any extra excitations (Fig. 4b,e) for either of the two circular polarizations. Nevertheless, there is an underlying effect arising from the interaction between the triskelia that causes the redshift of the main excitations as the separation between the triskelia increases. Consequently, the main features shown by the extinction CS (see Fig. 4c,f) as the edge-to-edge distance increases are driven by those of the absorption CS. Similar dependencies of the main and extra excitations on the edge-to-edge distance are found at 0° (not shown), but with the extra excitations appearing under both circular polarizations and larger values of the splitting between the main and extra excitations due to the lower twist angle.

A quick inspection of the CD of the absorption and scattering CS (Fig. 4g,h) reveals that the CD in the extinction signal arises mainly due to the two extra excitations found in the absorption CS under LCP light only (see Fig. 4i). In addition, and as discussed before, there are also small differences between the intensities of the two main excitations under LCP and RCP illuminations that yield some dichroism in the absorption and scattering CS as well, but almost cancelling each other out in the CD of the extinction CS due to the opposite sign of these two contributions as they are associated with less interacting modes of the two triskelia in the stack that are similar to those of a single triskelion. Thus, the extra excitations present in the absorption under LCP light give rise to a distinctive CD in the absorption CS without counterpart in the scattering CS that in turn causes most of the CD in the extinction CS.

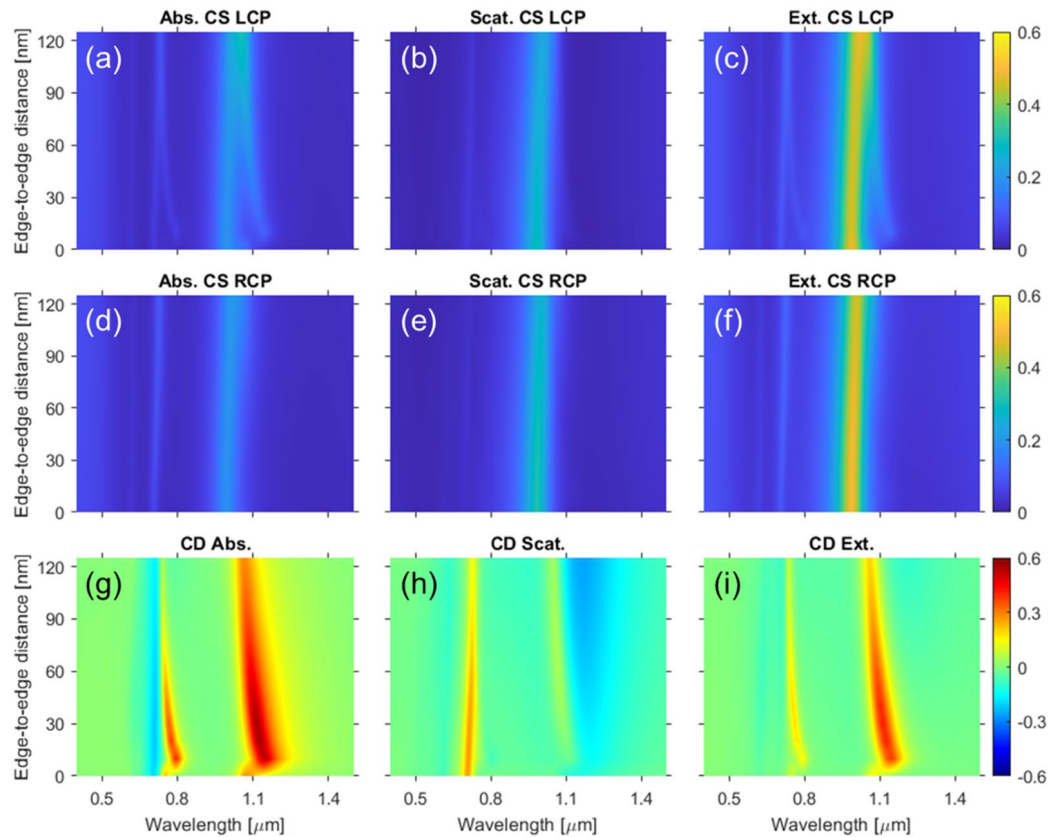

**Figure 4.** Absorption, scattering, and extinction CS for LCP (**a**–**c**), and RCP (**d**–**f**), illuminations, respectively, as a function of the edge-to-edge distance for a fixed twist angle $\varphi = 30°$. FOM defined in Eq. (1) for the CD of the absorption (**g**), scattering (**h**), and extinction (**i**) as a function of the edge-to-edge distance.

Finally, it is worth stressing that the study of the optical response as a function of the twist angle $\varphi$ in Sect. 6 was performed at an edge-to-edge distance of 30 nm. At this value of the distance, there is a certain overlap in the absorption between the LCP extra excitations and the main ones (see Fig. 4a). It happens in such a way that the absolute difference between the LCP and RCP signals is large around the excitations, while the CD in the extinction CS is still high enough, as shown in Fig. 4i. Besides, 30 nm is the midpoint of the range within about 10 and 50 nm for which the CD associated with the extra peaks is maximum.

**Near-field distributions of the modulus of the electric field.** The near-field distributions of the modulus of the electric field (intensity) in the vicinity (2 nm above) of each triskelion are shown in Fig. 5 and Fig. S2 in the Supplementary Information. The edge-to-edge distance between the two triskelia is 30 nm and the twist angles $\varphi$ are 30° and 60° (Fig. 5), and 0° and 90° (Fig. S2 in the Supplementary Information), respectively. The excitation of the extra resonances causing the CD in the extinction discussed in the previous section can now be easily correlated with the near-field distribution around the triskelia in each case. For instance, when the angle is 30°, the two triskelia display high intensity of the electric field around the metal for LCP illumination, corresponding to the excitation of the extra modes. It is also possible to observe the signature of the top triskelion on the plane of the bottom one as a blurry electric-field distribution resembling the spatial arrangement of the triskelion on top. On the contrary, under RCP illumination the overall electric response of the system is several times smaller since no extra resonances are excited for this polarization. Interestingly, the system shows a higher degree of optical activity under incident light with opposite polarization to the helicity of the structure, as previously reported in the literature[43]. The electric-field distributions are similar for 90° (see Fig. S2 in the Supplementary Information) but swapping the images corresponding to LCP and RCP light with respect to the 30° case, similarly to Ref[7] always showing the highest near-field activity in the case in which the helicity of the stack is opposite to that of the incident light.

Alternatively, when considering the configurations yielding no CD at all (60° and 0° in Fig. 5 and Fig. S2 in the Supplementary Information, respectively), the intensity distributions are almost indistinguishable for the two light polarizations. The case for 0° is like that of a single element because the two triskelia are parallelly aligned so that there is no chirality associated with the screw-like arrangement of the elements. In addition, the top element has a shadowing effect of the impinging radiation over the bottom one, hampering its excitation (more evident for the 0° case in Fig. S2 in the Supplementary Information). Even though there is no CD for 60° in any of the three studied optical functions, the intensity distributions look qualitatively different from those for 0° because the two triskelia display almost the same excitation regardless of the light polarization (see Fig. 5). It is worth stressing that for twist angles smaller than 60° the stack is more active under LCP illumination, whereas for angles higher than 60° the converse applies. Therefore, for 60° the system cannot show any differences on its excitation under either LCP or RCP light due to symmetry reasons.

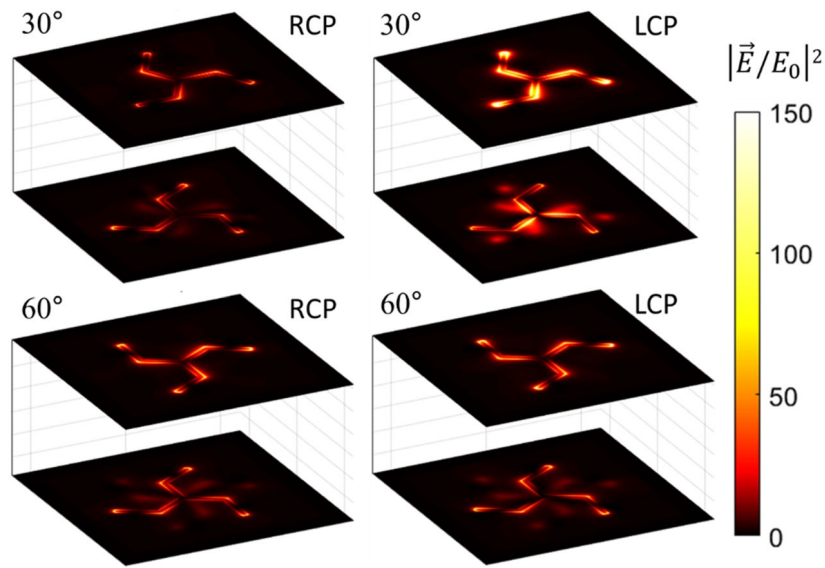

**Figure 5.** Near-field distributions of the square modulus of the electric field normalized to that of the incident light under RCP and LCP illumination at a wavelength of 1100 nm for twist angles $\varphi$ of 30° and 60°.

A deeper insight into the differences in the activity of the stack under LCP and RCP light for the extra excitations around 760 and 1100 nm (those mainly driven by interactions) can be gained by the near-field distributions of the modulus of the electric field in the middle plane between the two triskelia for $d = 30$ nm and $\varphi = 30°$ (see Fig. 6). As shown in Fig. 6b,d, under LCP light (opposite to the helicity of the stack), there is high intensity of the electric field between the two triskelia for both extra modes caused by the strong interactions between them, so that the near-field distributions have high 3D character and can be considered as excitations of the stack as a whole. Consequently, they give rise to CD in the extinction CS according to their true 3D nature. On the contrary, the stack under RCP light (see Fig. 6a,c) shows much lower activity in between the two triskelia as the excitations correspond to more independent modes of the two triskelia (more planar character), and, consequently, they do not significantly contribute to the CD in the extinction CS.

All things considered, following the results presented in previous sections, the study of the near-field distributions in the system reveals big differences in the intensity of the evanescent field under the two polarizations of light, providing a further proof of its chiral response.

## Conclusions

To summarize, the CD in the extinction CS exhibited by the twisted stack of triskelia is determined by the arrangement of the two elements. The extra resonances yielding the dichroic signal mostly appear in the absorption CS and are not intrinsic to a single triskelion. Most of the previous work found in the literature exploits and manipulates resonances already shown by the single element used to build the chiral structure. On the contrary, our work shows that the excitations responsible for the CD in the extinction CS are arising from the interaction between the two triskelia and are not present in the single triskelion case. Moreover, they are clearly interaction-driven since they show strong tunability by adjusting both the distance between the elements and their twist angle.

In addition, the helicity of the system strongly favors the excitation of the extra resonances under the illumination of one of the circular polarizations only (that opposite to the helicity of the structure) yielding near-field distributions between the two triskelia with 3D character that cause strong CD in the extinction CS. Special cases are those for 0° and 120°, for which the extra resonances show the same intensity for both RCP and LCP light (no helicity of the stack) and their contributions cancel each other in the extinction CS. As the angle increases/decreases from these values, the intensity for one polarization raises while rapidly vanishes for the opposite polarization. Thus, at an angle of 10° (110°) the extra resonances are only noticeable for one of the polarizations.

Finally, we have shown that by finely tuning the relative distance between the two triskelia and the twist angle of the structure, total control over the CD in the extinction can be achieved, obtaining tunable values up to 60% in the visible to near-infrared range. The simplicity with which these parameters can be manipulated and finely defined through successive lithographic and thin film deposition processes makes our design of special interest for the realization of more reliable and efficient chiral plasmonic nanostructures.

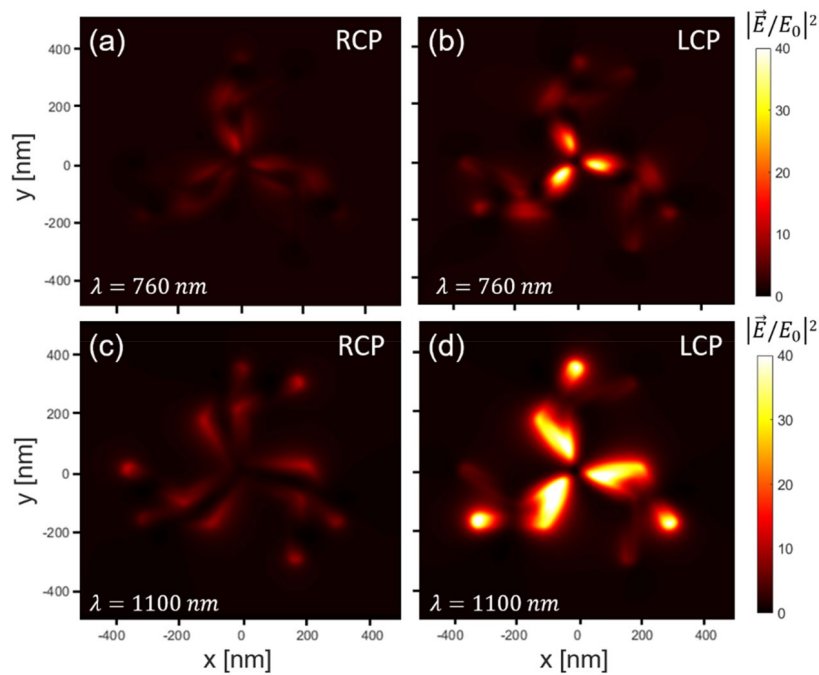

**Figure 6.** Near-field distributions of the square modulus of the electric field normalized to that of the incident light under RCP and LCP illumination at wavelengths of 760 nm (**a**, **b**), and 1100 nm (**c**, **d**), for twist angles $\varphi$ of 30° in a plane perpendicular to the axis of the stacking and equidistant to both triskelia.

This work constitutes a solid contribution to the polarization-dependent manipulation of light and holds promise for applications such as highly efficient helicity-sensitive surface spectroscopies or single-photon polarization detectors.

## Author contributions

J.R.-A. and A.G.-M. performed the simulations. J.R.-A, A.G.-M. and A.L. wrote the first version of the manuscript. All the authors participated in the discussion and review of the manuscript.

## Competing interests

The authors declare no competing interests.

## Additional information

**Supplementary Information** The online version contains supplementary material available at https://doi.org/10.1038/s41598-021-03908-2.

**Correspondence** and requests for materials should be addressed to J.R.

**Reprints and permissions information** is available at www.nature.com/reprints.

**Publisher's note** Springer Nature remains neutral with regard to jurisdictional claims in published maps and institutional affiliations.